\documentclass[a4paper,fleqn,usenatbib]{mnras}

\usepackage{mathptmx}

\usepackage[T1]{fontenc}
\usepackage{ae,aecompl}

\usepackage{graphicx}	
\usepackage{amsmath}	
\usepackage{amssymb}	
\usepackage{times,graphicx,amsmath,amsfonts,amssymb,epstopdf}
\usepackage[normalem]{ulem}

\newcommand{\msun}{{\rm M_\odot}/h}
\newcommand{\mpc}{h^{-1}{\rm Mpc}}

\title[Halo environment and screening in $f(R)$ gravity]
{Environmental screening of dark matter haloes in $f(R)$ gravity}

\author[D. Shi, B. Li and J. Han]{
Difu Shi$^{1}$\thanks{difu.shi@durham.ac.uk},
Baojiu~Li$^{1}$
and Jiaxin Han$^{2,1}$
\\
$^1$Institute for Computational Cosmology, Department of Physics, Durham University, South Road, Durham DH1 3LE, UK\\
$^2$Kavli Institute for the Physics and Mathematics of the Universe (WPI), Tokyo Institutes for Advanced Study, The University of Tokyo, 5-1-5 Kashiwanoha,\\
\ \ Kashiwa-shi, Chiba, 277-8583, Japan
}

\date{Accepted XXX. Received YYY; in original form ZZZ}

\pubyear{2017}

\begin{document}
\label{firstpage}
\pagerange{\pageref{firstpage}--\pageref{lastpage}}
\maketitle

\begin{abstract}
In certain theories of modified gravity, Solar system constraints on deviations from general relativity (GR) are satisfied by virtue of a so-called screening mechanism, which enables the theory to revert to GR in regions where the matter density is high or the gravitational potential is deep. In the case of chameleon theories, the screening has two contributions -- self-screening, which is due to the mass of an object itself, and environmental screening, which is caused by the surrounding matter -- which are often entangled, with the second contribution being more crucial for less massive objects. A quantitative understanding of the effect of the environment on the screening can prove critical in observational tests of such theories using systems such as the Local Group and dwarf galaxies, for which the environment may be inferred in various ways. We use the high-resolution {\sc liminality} simulation of \citet{liminality} to test the fidelity of different definitions of environment. We find that, although the different ways to define environment in practice do not agree with one another perfectly, they can provide useful guidance, and cross checks about how well a dark matter halo is screened. In addition, the screening of subhaloes in dark matter haloes is primarily determined by the environment, with the subhalo mass playing a minor role, which means that lower-resolution simulations where subhaloes are not well resolved can still be useful for understanding the modification of gravity inside subhaloes.
\end{abstract}




\section{Introduction}

One class of models used to explain the mysterious accelerating cosmic expansion, without invoking the addition of an exotic dark energy component \citep{copeland2006}, assumes that the standard theory of gravity, general relativity (GR), breaks down on cosmological scales and needs to be modified. Such `modified gravity' theories \citep{mg_review1,mg_review2} have generated considerable interest in the cosmological community in recent years. Even though so far there has been no widely-accepted alternative to the standard GR+$\Lambda$ cold dark matter ($\Lambda$CDM) model, in which the cosmic acceleration is driven by a positive cosmological constant, the study of possible alternatives could shed light on a question to which an answer is long over-due: does GR hold on cosmological scales \citep{koyama_mg_review,mg_review3}? With a number of large cosmological surveys having finished, on going, kicking off and being planned \citep[e.g.,][]{cfhtlens,boss,des,euclid,desi,erosita,lsst}, it is optimistic that we will soon enter a new era of research in this field.

Usually, it is assumed that a single gravitational equation governs the behaviour of gravity in the whole classical regime, which covers a huge range from the smallest scales where gravity has ever been tested (sub-millimetre) to scales comparable to the observable Universe. Therefore, a model of gravity can be tested on various scales and must pass all these tests in order to be viable. Given that GR has been well-established in the Solar system or other systems of relatively small size, such as binary pulsars \citep{binary_pulsar} and black holes \citep{ligo}, the behaviour of modified gravity models is expect to make the transition from being GR-like on small scales to predicting a cosmic acceleration on large scales. One way to achieve this, which has been the topic of intensive study in recent years, is via a screening mechanism, which suggests an environmental dependence of gravity: in environments similar to the Solar system GR is recovered, while allowing the scope for deviations in environments beyond the reach of current gravity experiments. In such models, two atoms would feel different gravitational forces due to one another depending on whether they are on Earth or in a low-density region.

The presence of a screening mechanism, in certain classes of modified gravity theories, not only leads to theoretical challenges, such as highly nonlinear gravitational field equations which render linear perturbation analyses more or less useless, but also has very practical implications for the testing of such models. Using the example above, the mutual gravity of the two atoms depends on their location. In other words, it is possible that the accurate prediction of the behaviour of gravity on the smallest scales (e.g., in lab experiments) depends on the actual status (and knowledge) of the much larger scale environment, such as the hosting galaxy or whether or not the galaxy is in a group or cluster.

Although the term `environment' has been used extensively in the literature, its precise meaning varies substantially both in theory and in practice. First of all, when we stated above that screening happens in dense environments {\it similar} to that of the Solar system, the {\it similarity} can be in terms of local matter density \citep[as in the case of the symmetron model,][]{symmetron}, the local Newtonian potential \citep[as in the chameleon model,][]{chameleon,chameleon2}, or derivatives of the potential [as in the cases of the Galileon \citep{dgp,galileon1,galileon2} and K-mouflage \citep{kmouflage1,kmouflage2} models].

On top of that, in reality, environment is often defined in terms of what one has from simulations or observations. For example, in a simulation with particle data, it is natural to define the environment of a halo as the average matter density in a spherical region of a given size around the halo, but this will be difficult to reproduce in observations, where instead one can quantify the environment by looking at how many galaxies are neighbouring a given galaxy (having more neighbours is usually an indication that a galaxy is in a high-density environment). Alternatively, using weak lensing one can construct maps of gravitational potential, by which the environment can also be quantified. Although the different measurements of environment rely on different physical quantities -- density, potential or derivatives of the potential -- and are therefore presumably suitable for testing the different theoretical screening mechanisms as mentioned above, as we shall see, there is a good correlation among them (e.g., a high-density region often has deeper gravitational potential and larger derivatives of the potential). This, together with the fact that there are only a limited number of ways to measure environment in observations, suggests that pragmatically all definitions of environment should be tried to see how best to understand the screening of modified gravity.

In addition, for objects with extensive sizes, such as those encountered in cosmology, depending on what we look at, the objects themselves can be part of the environment. For example, if we consider a massive galaxy cluster which hosts a galaxy, then the cluster itself (excluding the galaxy) serves as part of the environment of the galaxy, along with the larger-scale environment in which cluster is embedded. It can sometimes be useful to distinguish between the self-screening and environmental screening of the cluster, with the former defined as the screening of modified gravity caused by the cluster assuming that it is embedded in a vacuum. In practice, definitions of environment do not always separate the two effects cleanly, as we shall see below.

The effect of environmental screening in modified gravity was investigated previously for the chameleon \citep{chameleon_screening} and symmetron \citep{symmetron_screening} models, but these studies used one particular definition of environment, and were based on relatively low-resolution $N$-body simulations. It is our purpose to further these studies in two ways:
(i) We will try other definitions of environment in addition to the one adopted in \citet{chameleon_screening} and \citet{symmetron_screening}, to study how robust the qualitative conclusions of these studies are to the various definitions. In particular, this will tell us whether the different ways to measure environment observationally can corroborate or complement each other;
(ii) Our study here is based on a higher-resolution simulation, which will enable us to resolve smaller dark matter haloes to study the screening in different parts (e.g., inner versus outer) of a halo, and to investigate the screening of subhaloes as well. The study of these low-mass objects will be useful for accurately understanding how gravity behaves in such systems, which have been suggested to provide the strongest astrophysical constraints on potential deviations from GR \citep[e.g.,][]{bvj2013}. The screening of modified gravity from these low-mass haloes also depends more sensitively and complicatedly on their environments.

The model studied here is a variant of $f(R)$ gravity \citep{fr1,fr2} as proposed by \citet{hu-sawicki}. With suitable model parameters, this model is a special case of the chameleon-type theory studied by \citet{chameleon_screening} and many other authors. Despite being a special case with a rather ad hoc form of the gravitational action, the model is quite representative in the sense that similar qualitative behaviours can be found in other variants of viable $f(R)$ models, or more generally chameleon models \citep[see, e.g.,][]{general_chameleon1,general_chameleon2}, or even symmetron or dilaton \citep{dilaton,general_dilaton} models. Therefore, it can be used as a test case to estimate constraints on certain possible of deviations from GR.

The layout of this paper is as follows: in \S\ref{sect:model_simulation} we briefly describe the model, the simulation and the way used to find haloes/subhaloes; in \S\ref{sect:environments} we briefly introduce the different definitions of environment to be tested, and have a look at their correlations with each other using the simulation; in \S\ref{sect:env_screening} we show the environmental screening effects using these different environment definitions and in \S\ref{sect:conclusion} we discuss the implications of our results and conclude.

Throughout the paper we use the convention that a subscript $_0$ (overbar) denotes the current (cosmic mean) value of a quantity. We use the unit  $c=1$ ($c$ is the speed of light) unless otherwise stated.

\section{Simulations of \lowercase{$f$}$(R)$ gravity}

\label{sect:model_simulation}

$f(R)$ gravity is the most well-studied modified gravity theory in the context of understanding the origin of the cosmic acceleration, and there is a large body of literature on various aspects of this model. In order to avoid unnecessary repetition, we shall restrain from devoting space to yet another introduction to it. Interested readers are referred to the review articles \citep[e.g.,][]{defelice_review,sotiriou_review} for full details, or to one of the research papers for shorter but still self-contained descriptions, e.g. \citet{liminality}, which not only concisely describes the essential ingredients of $f(R)$ gravity sufficient for understanding this paper, but also introduces the {\sc liminality} simulation which this work is based on.

The {\sc liminality} simulation is a dark matter only $N$-body simulation of a certain variant of the Hu-Sawicki $f(R)$ gravity model \citep{hu-sawicki}. It was run using the {\sc ecosmog} code \citep{ecosmog}, which itself is based on the publicly available $N$-body  code {\sc ramses} \citep{ramses}, but includes new modules and subroutines to solve the modified Einstein equations in $f(R)$ gravity. This is an effectively parallelised adaptive mesh refinement (AMR) code, which starts with a uniform grid (the domain grid) covering the cubic simulation box with $N^{1/3}_{\rm dc}$ cells on a side. If the effective particle number in a grid cell becomes greater than a pre-defined criterion ($N_{\rm ref}$), the cell is split into eight ``son'' cells, and in this way the code hierarchically achieves ever higher resolution in dense environments. Such high resolution is necessary to accurately trace the motion of particles and guarantee the accuracy of the fifth force solutions. The force resolution, denoted by $\epsilon_{\rm f}$, is taken as twice the size of the cell where a particle is physically located, and we quote $\epsilon_{\rm f}$ on the highest refinement level. The simulation and model parameters are summarized in Table \ref{table:simulations}.

\begin{table}
\caption{The physical and technical parameters of the {\sc liminality} simulation and its associated $\Lambda$CDM simulation (this is for comparison and was run with exactly the same initial condition and simulation specifications). $\epsilon_{\rm s}$ is the threshold value of the residual \citep[see][for a detailed discussion]{ecosmog} which marks the convergence of the scalar modified gravity solver. The refinement criterion $N_{\rm ref}$ (see main text) is an array which takes different values at different refinement levels, and $\sigma_8$ is for the $\Lambda$CDM model only -- it was used to generate the initial conditions; its value for $f(R)$ gravity is different but is irrelevant for the analyses in this paper.}
\begin{tabular}{@{}lll}
\hline\hline
Parameter & Physical meaning & Value \\
\hline
$\Omega_{\rm b}$  & present fractional baryon density & $0.046$ \\
$\Omega_{\rm m}$  & present fractional matter density & $0.281$ \\
$\Omega_{\Lambda}$ & $1.0-\Omega_m$ & $0.719$ \\
$h$ & $H_0/(100$~km~s$^{-1}$Mpc$^{-1})$ & $0.697$ \\
$\sigma_{8}$ & linear r.m.s. density fluctuation & $0.820$ \\
$n_s$ & index of primordial spectrum & $0.971$ \\
\hline
$f_{R0}$ & HS $f(R)$ parameter & $-1.0\times10^{-6}$ (F6) \\
\hline
$L_{\rm box}$ & size of simulation box & 64~$h^{-1}$Mpc\\
$N_{\rm p}$ & particle number of simulation & $512^3$\\
$m_{\rm p}$ & particle mass of simulation & $1.52\times 10^{8}h^{-1}{\rm M_{\odot}}$\\
$N_{\rm dc}$ & cell number in domain grid & $512^3$\\
$N_{\rm ref}$ & criterion for refinement & 3, 3, 3, 3, 4, 4, 4, 4...\\
$\epsilon_{\rm s}$ & scalar solver convergence criterion & $10^{-8}$ \\
$\epsilon_{\rm f}$ & simulation force resolution & 1.95~$h^{-1}$kpc\\
\hline
$N_{\rm snap}$ & number of output snapshots & $122$ \\
$z_{\rm ini}$ & initial redshift & $49.0$ \\
$z_{\rm final}$ & stopping redshift & $0.0$ \\
\hline
\end{tabular}
\label{table:simulations}
\end{table}

In $f(R)$ models, the strength of gravity is enhanced compared to GR, and the size of the enhancement depends on the local gravitational potential \citep[e.g.,][]{chameleon}, ranging from $0$ inside deep potential wells (where screening takes place) to a maximum of $1/3$ in regions with shallow potential. The maximally $1/3$ enhancement of gravity is a generic property of the models, regardless of the technical details (e.g., whether it is Hu-Sawicki or some other variant), the latter only affecting the transition between $0$ and $1/3$ (e.g., whether at a given spacetime position the modification to GR is screened or not). Following the convention used in the literature, we call the difference between the modified and standard GR gravitational forces the {\it fifth force}. If screening happens, the fifth force vanishes, and in unscreened regions it is an attractive force $1/3$ the strength of standard Newtonian gravity, giving a total force of $4/3$ the GR force.

The Hu-Sawicki $f(R)$ model studied is a particular case specified by a parameter $f_{R0}\equiv\left[{\rm d}f(\bar{R})/{\rm d}\bar{R}\right]_0=-10^{-6}$. This case is of particular interest here, since the deviations from GR it predicts on cosmological scales are still allowed by current observations \citep[see, e.g.,][for some of the latest cosmological constraints on $|f_{R0}|$; note that a smaller $|f_{R0}|$ indicates a weak deviation from GR]{cataneo-cluster,liu_pc}; while stronger constraints are suggested from smaller scales, a precise quantification of the constraints requires a good knowledge about whether (and how well) dark matter haloes and subhaloes are screened \citep{bvj2013}. In the literature, this model is often called F6; throughout the paper, when we talk about $f(R)$ model, we mean F6 unless clearly otherwise stated. With $512^3$ particles in a box of size $L_{\rm box}=64\ h^{-1}\rm Mpc$, the {\sc liminality} simulation is currently the highest resolution {cosmological} simulation of $f(R)$ gravity that runs from $z=49$ until $z=0$, with full information of the fifth force recorded. As a result, it is ideal for the analysis of the screening of dark matter haloes and their substructures (which would be poorly resolved should the resolution be too low).

The dark matter halo catalogue used in our analyses was obtained using the friends-of-friends (FoF) group-finding algorithm, with a linking length of $0.2$ times the mean inter-particle separation \citep{Davis1985}. We used the tracking subhalo finder Hierarchical Bound-Tracing \citep[][{\sc hbt}]{HBT} to identify subhaloes. {\sc hbt} works in the following way:
(i) starting from isolated haloes at a previous snapshot, it finds their descendants in subsequent snapshots and keeps track of their evolution; (ii) when two haloes merge, it tracks the self-bound part of the less massive progenitor as a subhalo in subsequent snapshots. In this way, all the subhaloes formed from halo mergers can be identified with a single walk through all the snapshots. This algorithm enables {\sc hbt} to largely avoid the resolution problem encountered by configuration-space subhalo finders.

\section{Environment definitions}

\label{sect:environments}

\begin{table}
\caption{The different definitions of environment used in this paper. See the main text for more details (\S\ref{sect:environments}).}
\begin{tabular}{@{}llll}
\hline\hline
name & symbol & parameter(s) & equation \\
\hline
conditional nearest neighbour & $D_{N,f}$ & $f=1$, $N=1,10$ & Eq.~(\ref{eq:D_N}) \\
spherical overdensity & $\delta_{R}$ & $R=5,8h^{-1}$Mpc & Eq.~(\ref{eq:Sph_ave}) \\
shell overdensity & $\delta_{R,R_{\rm min}}$ & $R_{\rm min}=R_{\rm halo}$ & Eq.~(\ref{eq:Shl_ave}) \\
experienced gravity & $\Phi_{\star}$ & None & Eq.~(\ref{eq:Phi}) \\
total gravity & $\Phi_{+}$ & None & Eq.~(\ref{eq:Phi}) \\
\hline
\end{tabular}
\label{table:environments}
\end{table}

As directly measuring the distribution of mass is not always possible, observers usually use the distribution of galaxies to estimate the density around galaxies. There are quite a few different methods to estimate the environmental dependence of galaxy properties. Table 1 in \citet{Haas2012} briefly summarized the environmental measures used in the literature. In simulations, using similar environmental measure makes it convenient to compare with observational results. On the other hand, the density field can be directly measured using simulation particles. These quantities are in principle more accurate than indirect environmental measures.

In this section, we briefly introduce the different definitions of environment we use.

\subsection{Conditional nearest neighbour}

Galaxies that live in denser environments preferentially have closer neighbours. Following this principle, the conditional nearest neighbour environment measure of a halo with mass $M_{\rm L}$ is defined as the distance $d$ to its $N$th nearest neighbour halo whose mass is at least $f$ times as large as $M_{\rm L}$ \citep{Haas2012}. This quantity is rescaled by the virialized radius $r_{\rm NB}$ of the neighbouring halo to define $D_{N,f}$ as
\begin{equation}
    D_{N,f}=\frac{d_{N,M_{\rm NB}/M_L\ge f}}{r_{\rm NB}}.
	\label{eq:D_N}
\end{equation}
A halo with large value of $D_{N,f}$ indicates a paucity of nearby haloes, implying that the halo lives in a low-density environment. Here, we only consider isolated haloes in analysis, i.e. subhaloes are not regarded as neighbour.

In the context of modified gravity, this environment definition was previously used in \citet{chameleon_screening} and \citet{symmetron_screening}. It has the added freedom of varying the values of $N$ and $f$ to allow continuous quantitative changes in $D_{N,f}$. It is also more directly connected to observations, as we can treat haloes and subhaloes as proxies to clusters and galaxies in the real Universe.

$D_{N,f}$ is also a faithful definition of `environment', because the halo of mass $M_{\rm L}$ itself is not counted as a neighbour (in other words, $D_{N,f}\neq0$). However, this definition is not completely independent of $M_{\rm L}$, as $M_{\rm L}$ is used in the condition $M_{\rm NB}/M_{\rm L}\geq f$ (which is why we name it the {\it conditional} nearest neighbour). This could lead to the unphysical consequence that for very massive haloes, which are likely to live in dense environments, it is more difficult to find neighbours with $M_{\rm NB}\geq fM_{\rm L}$. Hence $D_{N,f}$ is large, falsely implying that such haloes are in low-density environments. We shall bear this in mind when analysing our results.

\subsection{Spherical \& shell overdensity}

In observations, counting the number of neighbouring galaxies in a fixed volume around a galaxy is another way to measure the environment, as a higher galaxy number density indicates a denser environment. Although galaxies are biased tracers of the underlying matter density field, techniques have been developed to reconstruct the density field from observational distributions of galaxies \citep[e.g.,][]{kitaura2010,platen2011,ata2016}.


Given our purely theoretical interest, we measure the dark matter density in a spherical volume around a halo, which we define as the spherical overdensity environment. This is expressed as
\begin{equation}
    1+\delta_R\equiv\frac{\rho(\leq R)}{\bar{\rho}}=\frac{N(\leq R)}{\bar{N}},
	\label{eq:Sph_ave}
\end{equation}
where the $R$ is the radius (in units of $h^{-1}$Mpc) of the spherical volume, $N$ is the number of particles found in this volume and $\bar{N}$ is the mean number of dark matter particles in a volume of size $4\pi R^3/3$.

By definition, the spherical overdensity environment measure $\delta_R$ includes the contribution from the halo at the centre of the spherical volume. One can define similarly a `shell overdensity' environment as
\begin{equation}
    1+\delta_{R,R_{\rm min}}\equiv\frac{\rho(R_{\rm min}\leq r\leq R)}{\bar{\rho}}=\frac{N(R_{\rm min}\leq r\leq R)}{\bar{N}},
	\label{eq:Shl_ave}
\end{equation}
where we exclude the particles within a minimum radius $R_{\rm min}$ given by $R_{\rm halo}\leq R_{\rm min}<R$, with $R_{\rm halo}$ being the radius of the central halo.

\begin{figure*}
	\includegraphics[width=\textwidth]{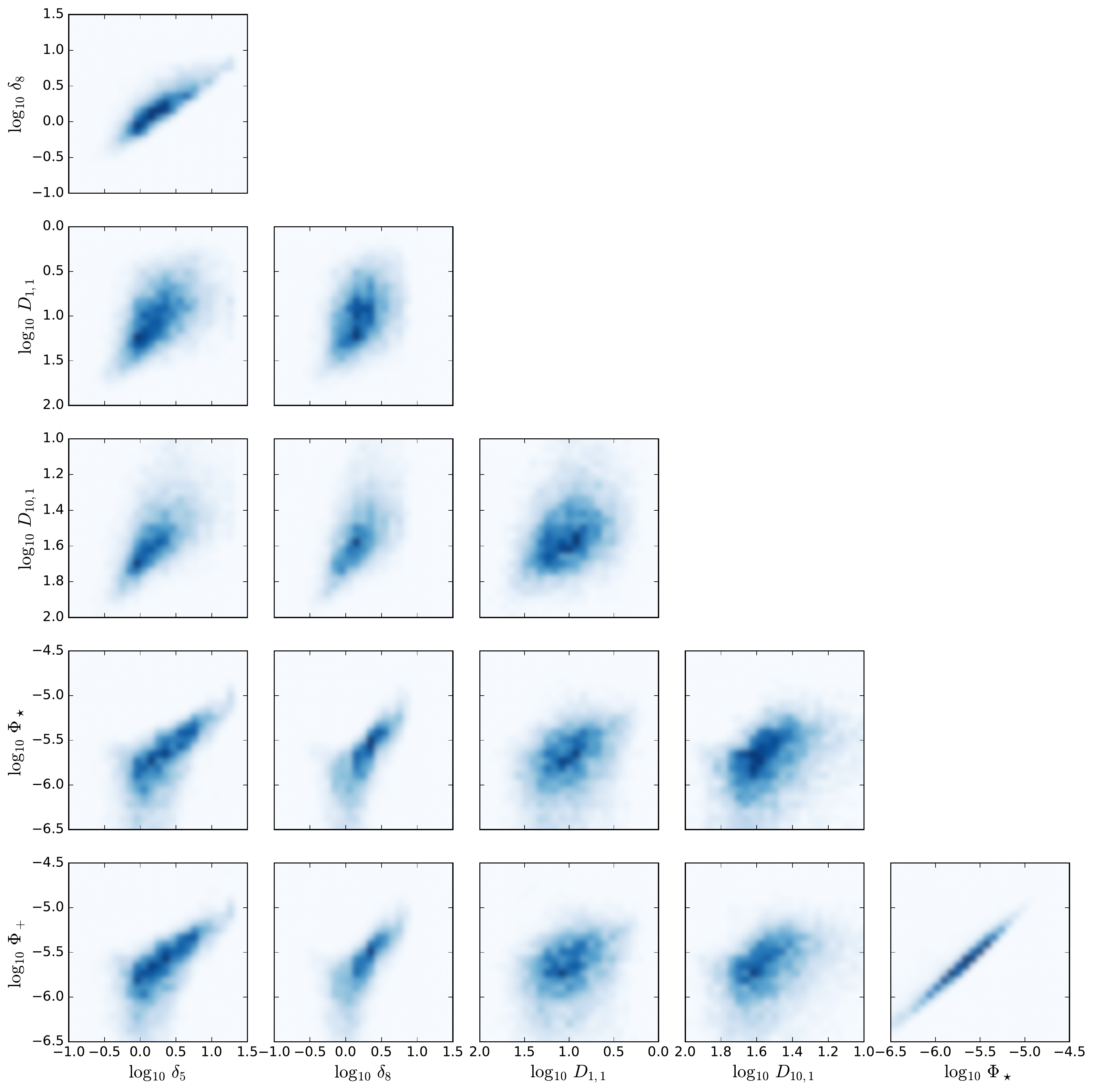}
    \caption{The correlations between the different environment measures defined in \S\ref{sect:environments}. The different panels compare different pairs of environment definitions, whose values are shown on the two axes of the panel. The number density of haloes in a given environment is colour coded, with darker (lighter) blue meaning more (fewer) haloes having that environment. The results are obtained using the {\sc liminality} simulation (Table \ref{table:simulations}). Note that due to space limitations not all environments used in the latter part of this paper are compared here.}
    \label{fig:corr}
\end{figure*}

\subsection{Experienced gravity}


While the dark matter density as used in defining $\delta_R$ and $\delta_{R,R_{\rm min}}$ is not directly measurable, its effects can be observed in various ways, such as gravitational lensing and galactic dynamics, that probe the lensing and dynamical potential respectively. Even though the two potentials may not coincide with each other in theories of modified gravity, they both serve as a good characterization of environment. Indeed, lensing and galaxy dynamics are governed respectively by the potential and its derivative, so using these to define the environment may be particularly useful for models in which the screening depends on these quantities.

In our simulation, the potentials of the Newtonian gravity and the total gravitational force in $f(R)$ gravity on every simulation particle are outputted separately. Here, we use only the Newtonian potential as an environment measure, since in the $f(R)$ model the lensing potential (which can be reconstructed from lensing observations) satisfies the standard Poisson equation as the Newtonian potential in GR subject to corrections that are negligible in practice.

The Newtonian potential at any given position inside a dark matter halo receives contributions from both self gravity (i.e., the potential due to the halo itself) and environment (i.e., that caused by material outside the halo). The self-gravity contribution can be calculated analytically given that haloes satisfy the usual Navarro-Frenk-White \citep[][NFW]{NFW,NFW2} density profile even in $f(R)$ gravity \citep{lombriser2012,liminality}. Subtracting this from the total Newtonian potential at the position, which is given by our simulation, leads to the environment measure that we dub ``experienced gravity''.

To be explicit, the Newtonian potential of a spherical halo is given as
\begin{equation}
    \Phi(r)=\int_{0}^{r} \frac{GM(r')}{r'^2}{\rm d}r' +C,
	\label{eq:halo_grav}
\end{equation}
in which $GM(r)/r^2$ is the gravitational force at distance $r$ from the centre of the halo, and $C$ is an integration constant that can be fixed using $\Phi(r\rightarrow\infty)=\Phi_\infty$, the Newtonian potential infinitely far away from the halo.

The NFW density profile of a spherical halo is given by
\begin{equation}
    \frac{\rho(r)}{\rho_{\rm c}}=\frac{\beta}{\frac{r}{R_{\rm s}}\left(1+\frac{r}{R_{\rm s}}\right)^2},
	\label{eq:NFW_prof}
\end{equation}
where $\rho_c$ is the critical density for matter, $\beta$ and $R_s$ are two fitting parameters. Plugging this into Eq.~(\ref{eq:halo_grav}), it can be derived that
\begin{equation}
    \int_{0}^{r} \frac{GM(r')}{r'^2}\mathrm{d}r'=4\pi G \beta \rho_{\rm c} R_{\rm s}^3 \left[\frac{1}{R_{\rm s}}-\frac{\ln\left(1+\frac{r}{R_{\rm s}}\right)}{r} \right],
	\label{}
\end{equation}
and so
\begin{equation}
    C=\Phi_\infty-4\pi G \beta \rho_{\rm c} R_{\rm s}^2.
	\label{}
\end{equation}
Here, if halo is isolated, then $\Phi_\infty=0$. But in $N$-body simulations or the real Universe, no halo is totally isolated from the others. So $\Phi_\infty$ does not always go to zero. Therefore, we replace $\Phi_\infty$ by $\Phi_\star$, which is the potential produced by all the other haloes at the position of a given halo. The Newtonian potential in the halo can then be written as
\begin{equation}
    \Phi(r)=\Phi_\star-4\pi G \beta \rho_{\rm c} \frac{R_{\rm s}^3}{r} \ln \left(1+\frac{r}{R_{\rm s}} \right),
	\label{eq:Phi}
\end{equation}
in which $\Phi_\star$ is our definition of the experienced gravity environment measure, and $\Phi(r)$ is directly measured from our simulation. The second term on the right-hand side of Eq.~(\ref{eq:Phi}) is the self-gravity contribution which is calculated from the NFW fitting parameters (i.e. $\beta, R_{\rm s}$) of every individual dark matter halo, cf.~Eq.~(\ref{eq:NFW_prof}). This way to compute $\Phi_\star$ was previously used in, e.g., \citet{lmb2011}.

Assuming that the size of a given halo is much smaller than the Universe or the simulation box, we would expect the environment measure $\Phi_\star$ to stay roughly constant inside a halo. In Eq.~(\ref{eq:Phi}), both $\Phi(r)$ and the self-gravity term on the right-hand side take different values at different $r$ ($r\leq R_{\rm halo}$), which means that there is no {\it a priori} guarantee that $\Phi_\star$ is the same everywhere at $r\leq R_{\rm halo}$. In Appendix \ref{sect:appendix_a}, we perform a check of the constancy of $\Phi_\star$ and show that the $\Phi_\star$ environment measure works quite well.

Finally, we note that $\Phi_\star$, although a good measurement of the environment of a dark matter halo, is not what gravitational lensing reconstructions give us as the latter do not distinguish between self and environmental contributions. For this reason we define another measure, called total gravity, or $\Phi_+$, which is the average of $\Phi(r)$, cf.~Eq.~(\ref{eq:Phi}), inside the halo ($r\leq R_{\rm halo}$). Neither $\Phi_\star$ nor $\Phi_+$ have extra free parameters, unlike $D_{N,f}$, $\delta_R$ and $\delta_{R,R_{\rm min}}$.

Table \ref{table:environments} summarizes our environment measures. Before looking at how the screening depends on environment, in Fig.~\ref{fig:corr} we first have a look at the correlation between the different environment measures themselves. Note that a strong correlation is present between the spherical overdensity ($\delta$) and experienced gravity ($\Phi$) definitions, which is as expected given the relationship between the gravitational potential and local matter density. In particular, $\Phi_\star$ has an extremely tight correlation to $\Phi_+$. As our test, for small haloes, their self-gravity is negligible to the environment; for massive haloes, the correlation is actually scattered. However, due to the very small number density of massive haloes, it is hard to see this scatter in Fig.~\ref{fig:corr}.
The conditional nearest neighbour measure ($D$), on the other hand, correlates much less tightly with the other measures. Even $D_{1,1}$ and $D_{10,1}$ are barely correlated to each other. The reason is the complexities in inferring the matter density from galaxy number density (or in our case, the halo density) due to the unknown bias.




Overall, the conclusion is that the different environment measures tested here do correlate reasonably well with one another.

\section{Environment measures and screening}

\label{sect:env_screening}

We use $M_{200}$ as halo mass, which is the mass inside the radius $r_{200}$ within which the average density is 200 times the critical density, $\rho_c$. In our analysis, dark matter haloes are binned into four mass ranges as $1\sim3\times 10^{11}\msun,\ 3\times 10^{11}\sim 1\times 10^{12}\msun,\ 1\times 10^{12}\sim1\times 10^{13}\msun$ and $1\times 10^{13}\sim1\times 10^{14}\msun$. Note that the $10^{11}\sim10^{12}\msun$ haloes are divided into two smaller bins. As the effects of modified gravity are efficiently screened in massive haloes, the main difference between F6 and GR is in these low-mass haloes.

Here, we study how the modified gravitational force and potential behave in different environments, for the environment measures introduced above. To this end, we define the {\it fifth-force-to-gravity ratio} (or {\it fifth force ratio} in short) as the ratio between the magnitude of the fifth force in $f(R)$ gravity (see~\S\ref{sect:model_simulation}) and that of the standard Newtonian force. This quantity approaches $0$ in screened regions and $1/3$ in unscreened regions, but can take any value in between (the transition region). We will also use the {\it fifth force potential} that is expressed in units of $2\Phi_{\rm N}/3$, where $\Phi_{\rm N}$ is the potential for the standard Newtonian gravity \citep{zhao2011}. Note that the fifth force potential is dimensionless, going to 0 in screened regions and $1/2$ in unscreened regions.

Our simulation outputs the fifth force and potential at the positions of all simulation particles. We measure the fifth force ratio and fifth force potential in two ways: the average over the values at the positions of all particles inside $R_{\rm halo}=R_{200}$, and the value at halo centres, the latter being obtained by averaging all the particles within $r\leq0.2R_{200}$ given the uncertainty in defining the halo centre, and call these $R_{c200}$ and `halo centre' respectively. The second case is relevant for the screening of the fifth force inside central galaxies that are at the centres of their host haloes, while the first case can be used for satellite galaxies.

\begin{figure*}
\includegraphics[width=\textwidth]{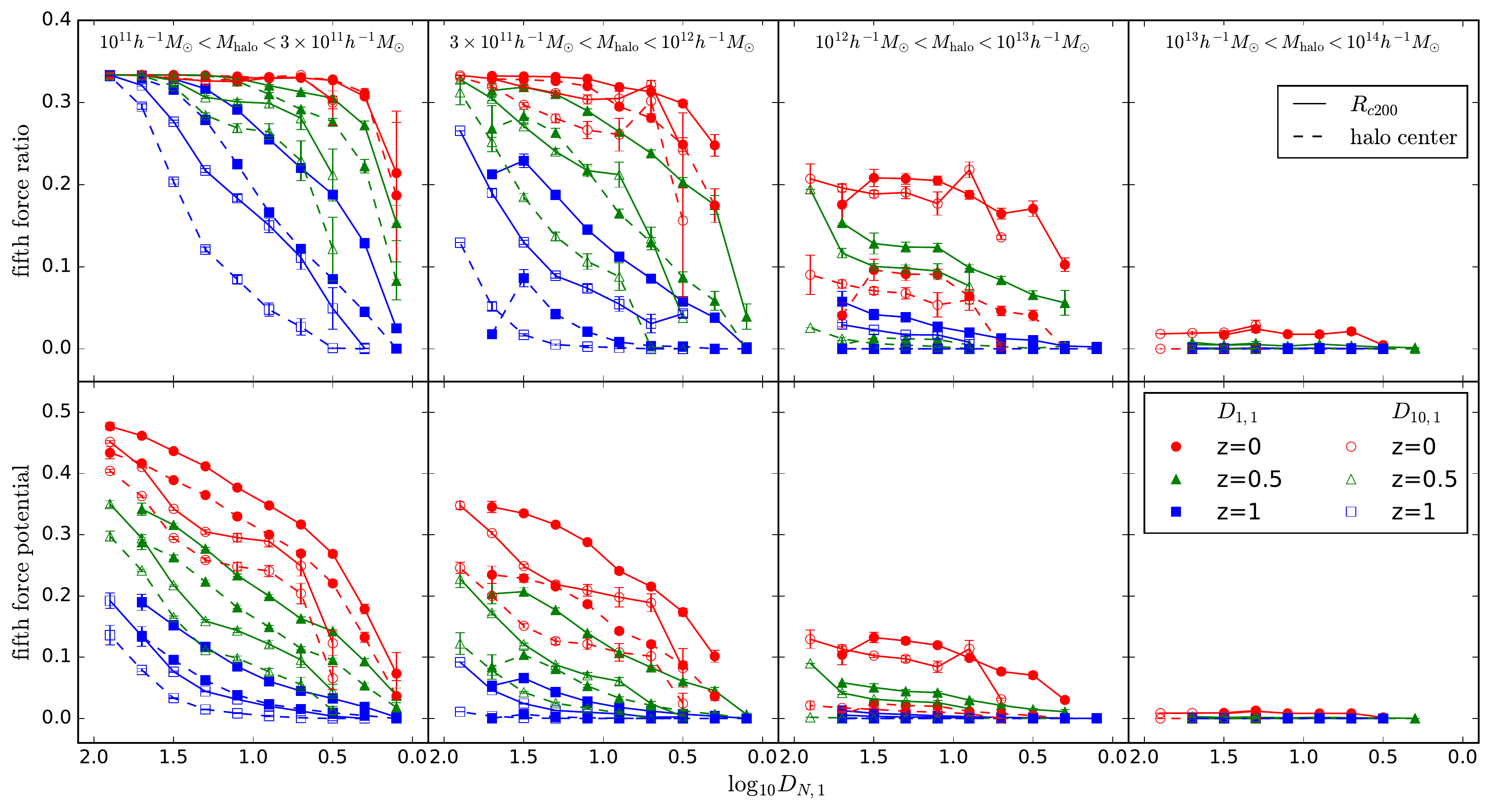}
\caption{The fifth force ratio (top panels) and fifth force potential (bottom panels) as a function of the conditional nearest neighbour halo environment $D_{1,1}$ (filled symbols) and $D_{10,1}$ (open symbols) at $z=0$ (red), $0.5$ (green) and $1$ (blue). The halo samples are divided in to four mass bins as indicated on the top of each panel. The solid lines show the results measured from all particles within $r_{200}$, and the dashed lines are measured from the halo centres only (see the text for more details). }
\label{fig:Neighbour}
\end{figure*}

\subsection{Conditional nearest neighbour}

In GR, $D_{1,1}$ is known to represent the local dark matter density well and to be almost uncorrelated with the mass of the halo. In $f(R)$ gravity, \citet{chameleon_screening} confirmed that the mass independence of $D_{1,1}$ still holds. We adopt both $D_{1,1}$ and $D_{10,1}$, which are derived from the first and tenth nearest neighbours heavier than the halo, respectively, as the conditional nearest neighbour environment definitions.

Fig~\ref{fig:Neighbour} shows the fifth force ratio (upper panels) and the fifth force potential (lower panels) against these two environmental measures, in the four halo mass bins (as indicated on the top of the different panels) at $z=0$ (red circles), $0.7$ (green triangles) and $1$(blue squares). At all three redshifts and in all four mass bins, there is a noticeable trend that the fifth force ratio and potential both increase with $D$. As larger values of $D$ correspond to lower density environments, this agrees with the expectation that the haloes living in low-density regions are more likely to be unscreened. Also, we can see clearly that the most massive haloes have negligible fifth force potentials, because of the efficient screening in these objects. The increase of fifth force ratio from $0$ to $1/3$ represents the transition from unscreened to screened haloes. At the same redshift, the transition occurs at smaller $D$ for lower mass haloes. Inside the same halo mass bin, the screening is stronger at higher redshifts, because the Universe is denser at early times.

The halo centre (dashed curve) always has a smaller fifth force ratio and potential than the average in the whole halo (solid curve). This is as expected, as the NFW profile has higher density in the inner region of a halo than at its outskirts, which means that screening is stronger in the inner part. The difference between halo centre and $R_{\rm c200}$ is particularly strong for the halo mass bin $10^{12}\sim10^{13}\msun$ at $z<0.5$, which is also true for less massive haloes at higher redshifts ($0.5<z<1$; blue and green curves in the two left columns). This is again because at a given redshift the fifth force inside haloes with a certain mass goes through a transition from screened to unscreened, and these haloes can be in a state such that their inner parts are well screened while the outer regions remain unscreened. This transition starts from smaller haloes first, and progressively affects more massive haloes at later times. This observation is relevant if one is interested in the screening of central galaxies in haloes.

Since $D_{N,f}$ depends on both the separation to the neighbour and the size of the neighbour, $D_{1,1}$ does not have to be smaller than $D_{10,1}$. Statistically, however, because the tenth nearest larger neighbouring halo of a given halo is always farther away than the nearest larger neighbour, and yet is not necessarily larger in size, $D_{1,1}$ is generally smaller than $D_{10,1}$. We can see in Fig.~\ref{fig:Neighbour}, for the same value of fifth force ratio/potential, $D_{10,1}$ is always larger than $D_{1,1}$. The $D_{10,1}$ curves shift to larger values along the $x$-axis compared to the $D_{1,1}$ curves. Conversely, $D_{1,1}$ represents a less dense environment than the same value of $D_{10,1}$ does. In Fig.~\ref{fig:Neighbour}, the fifth force ratio curves corresponding to the $D_{10,1}$ environment measure (open symbols) are consistently below those for $D_{1,1}$ (filled symbols), confirming that screening is stronger for denser environments.

\subsection{Spherical overdensity}

\begin{figure*}
\includegraphics[width=\textwidth]{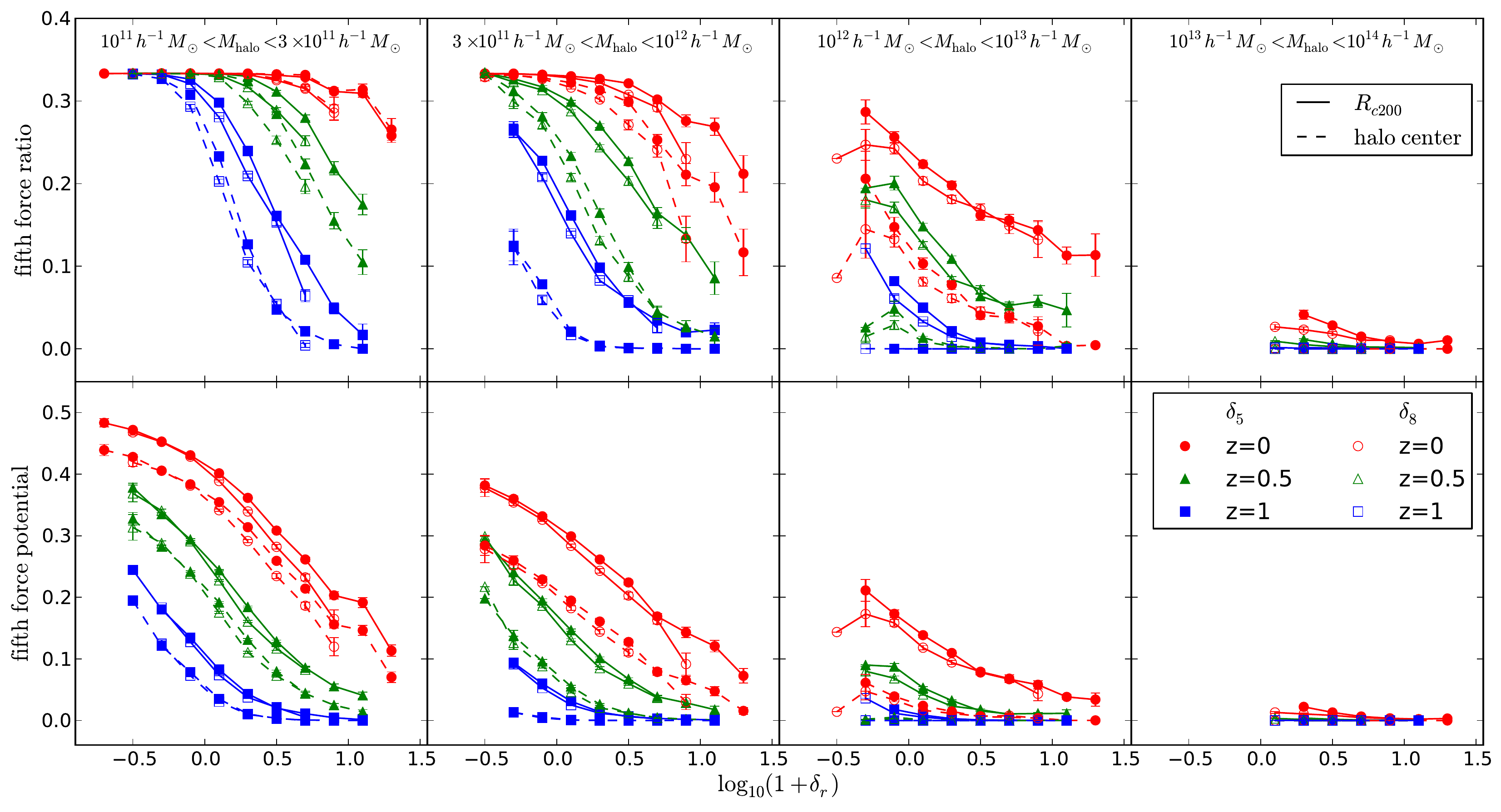}
\caption{The fifth force ratio (top panels) and fifth force potential (bottom panels) as a function of the spherical overdensity halo environment $\delta_5$ (filled symbols) and $\delta_{8}$ (open symbols) at $z=0$ (red), $0.5$ (green) and $1$ (blue). The halo samples are divided in to four mass bins as indicated on the top of each panel. The solid lines show the results measured from all particles within $R_{200}$, and the dashed lines are measured from the halo centres only (see the text for more details).}
\label{fig:sph_avg}
\end{figure*}

We use spherical volumes with radii of $5$ and $8\mpc$ around the dark matter halo centre to measure the density contrast, devoted $\delta_{5}$ and $\delta_8$. These values are used in \citet{halo_void} to ensure the spherical volumes are neither too large (otherwise they will not be a faithful representation of the local environment) nor too small (otherwise the definition of environment will be too noisy and sensitive to the presence of the central halo in the spherical volume). Physically, these numbers are up to a few times the Compton wavelength of the scalar field for the redshift range we are interested in, and the region enclosed is most relevant to the dynamical state of the scalar field\footnote{To be more explicit, if the region is too small, e.g., smaller than the scalar field's Compton wavelength, then the scalar field will be affected by matter outside, making the environment definition insufficient. Similarly, if the spherical region is too large, it may contain matter which does not have a significant impact on the scalar field in the central halo, again making the definition less relevant.}.

Fig.~\ref{fig:sph_avg} plots the spherical overdensity measures against the fifth force ratio and potential, which has the same format as Fig.~\ref{fig:Neighbour}. We can see clearly the similar overall trend that the fifth force ratio and potential decrease with increasing $\delta_R$. Here, a larger value of $\delta_R$ means higher density region, which in turn means the fifth force is more likely to be screened, again, as expected.

For the same value of $\delta_5$ and $\delta_8$, $\delta_8$ represents an environment with higher matter density because it manages to have the same $\delta$ in a bigger volume even though matter over-density is generally expected to be lower at larger radii from the centre. Correspondingly, Fig.~\ref{fig:sph_avg} shows that the screening is stronger for $\delta_8$. However, the difference is small, because even $5\mpc$ is already significantly bigger than the halo radius. 

Comparing Fig.~\ref{fig:sph_avg} to Fig.~\ref{fig:Neighbour}, it can be seen that these two plots are qualitatively similar to each other. This serves as a cross check that both environment definitions can be applied to infer the screening of galaxies for an observed galaxy catalogue. We also checked the results by using the shell overdensity definition, $\delta_{R,R_{\rm min}}$, where $R_{\rm min}=R_{\rm halo}=R_{200}$ and $R=5\mpc, 8\mpc$, and found only tiny differences from Fig.~\ref{fig:sph_avg}; so we will not show them here.


\subsection{Experienced and total gravity}

Finally, we consider the experienced and total gravity measures of halo environment, which are defined using the Newtonian potential produced at the position of a halo by matter outside the halo and by all matter (including that from the halo itself) respectively.

In order to fit the NFW profile, we divide the halo radius, $R_{200}$, into $20$ bins equally spaced in logarithmic scale [see \citet{liminality} for more details], and then measure the mass density of every spherical shell. $\Phi_\star$ is then calculated using Eq.~(\ref{eq:Phi}) and the NFW parameters resulting from the fit, in which $\Phi(r)$ is read from the simulation output and spherically averaged for every shell. The $\Phi_\star$ calculated in this way has small fluctuations across different shells, due to the detailed mass distribution in the halo, and due to numerical noise, but in Appendix \ref{sect:appendix_a} we can see that the fluctuations are insignificant. The value of $\Phi_\star$ used is the average over all spherical shells.

\begin{figure*}
\includegraphics[width=\textwidth]{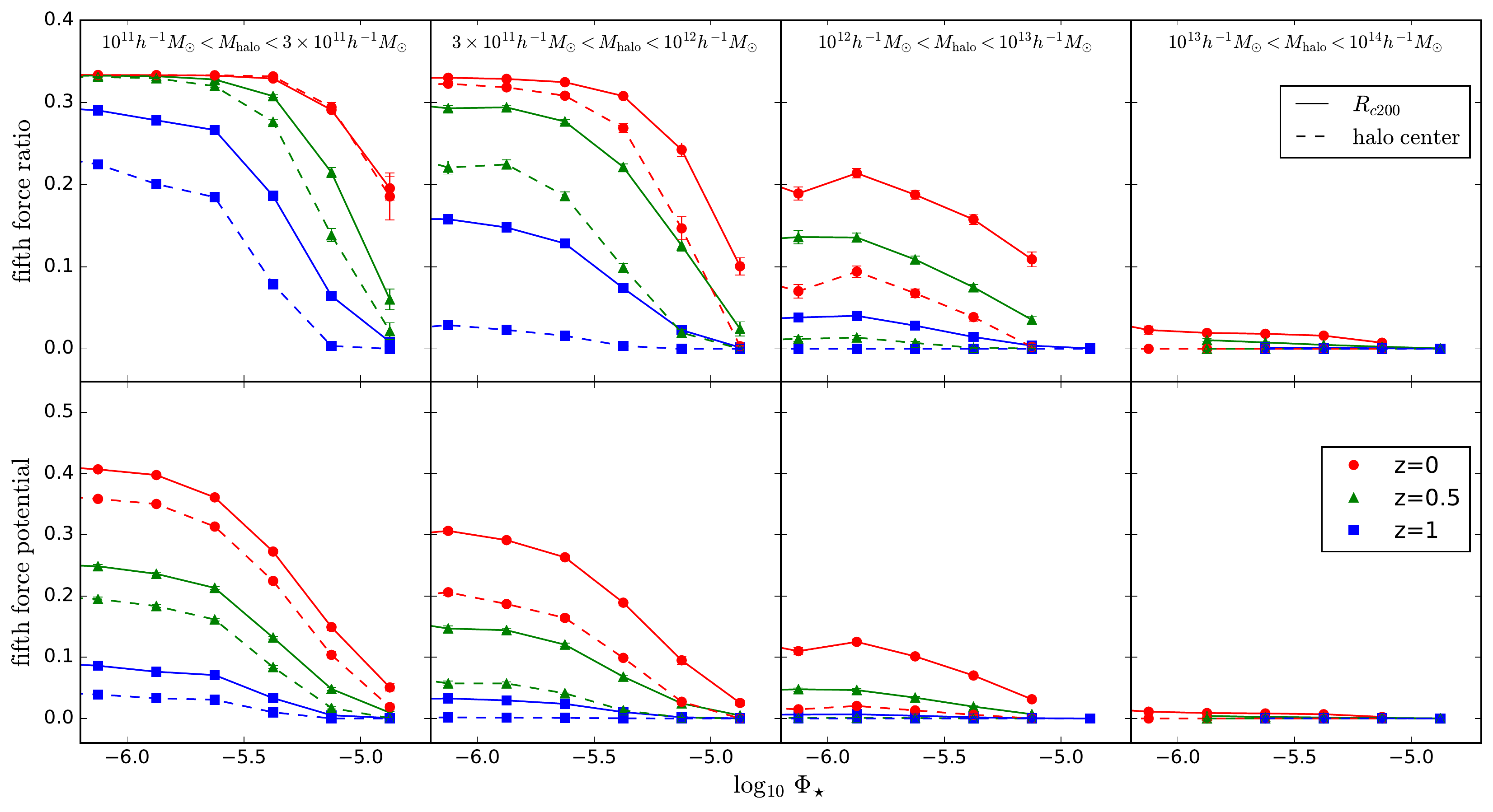}
\caption{The fifth force ratio (top panels) and fifth force potential (bottom panels) as a function of the experienced gravity halo environment $\Phi_\star$ (filled symbols) at $z=0$ (red), $0.5$ (green) and $1$ (blue). The halo samples are divided into four mass bins as indicated on the top of each panel. The solid lines show the results measured from all particles within $R_{200}$, and the dashed lines are measured from the halo centres only (see the text for more details).}
\label{fig:Phi}
\end{figure*}

Fig.~\ref{fig:Phi} shows how the fifth force ratio and fifth force potential depend on $\Phi_\star$ for the four mass bins considered. We can see a similar overall trend to that in Figs.~\ref{fig:Neighbour} and \ref{fig:sph_avg}, that more massive haloes living in denser environments (i.e., larger $|\Phi_\star|$) are better screened, but there is also a noticeable difference, namely the curves in Fig.~\ref{fig:Phi} are smoother and the scatter smaller. The latter, in particular, implies that this definition of environment is more reliable and less sensitive to the unknown bias of observed galaxies for the study of the chameleon screening. This is not unexpected, since it is well known that the condition for screening -- the thin-shell condition \citep{chameleon} -- is explicitly determined by the Newtonian potential that an object feels.

\begin{figure*}
\includegraphics[width=\textwidth]{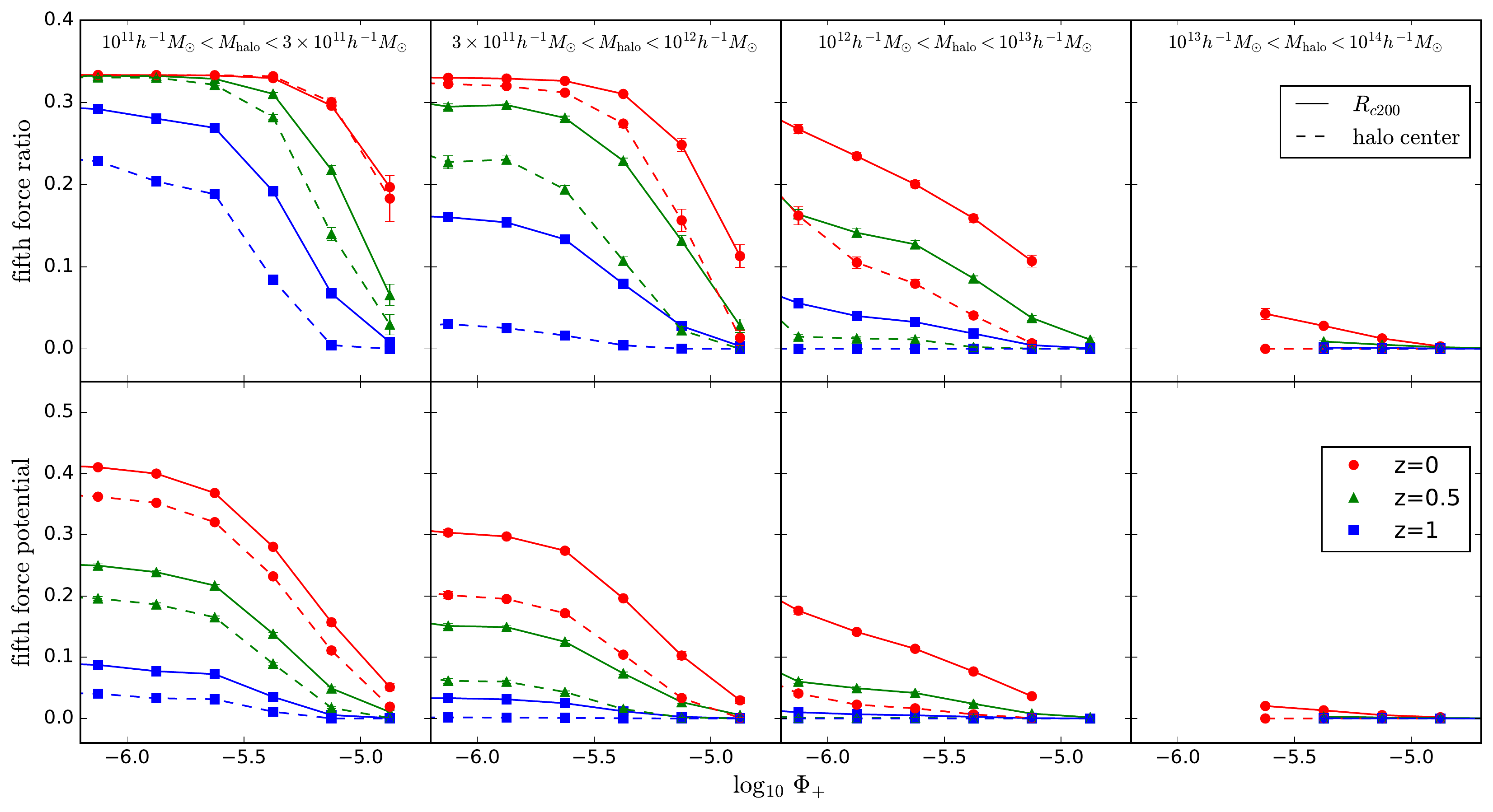}
\caption{The fifth force ratio (top panels) and fifth force potential (bottom panels) as a function of the total gravity potential halo environment $\Phi_+$ (filled symbols) at $z=0$ (red), $0.5$ (green) and $1$ (blue). The halo samples are divided into four mass bins as indicated on the top of each panel. The solid lines show the results measured from all particles within $R_{200}$, and the dashed lines are measured from the halo centres only (see the text for more details).}
\label{fig:Phi_plus}
\end{figure*}

In Fig.~\ref{fig:Phi_plus} we present the same result as in Fig.~\ref{fig:Phi}, but with $\Phi_\star$ replaced by $\Phi_+$. All features discussed in Fig.~\ref{fig:Phi} remain, with only slight quantitative changes. In particular, the curves are smooth and the scatter is small. As $\Phi_+$ is the total potential at the position of a dark matter halo, it is what weak lensing (tomography) observations will give us; this is unlike $\Phi_\star$, which is the potential at the position of the halo produced by everything but the halo itself, and thus is a more theoretical definition of `environment'.

\section{Environmental screening of subhaloes}


With the current knowledge of galaxy formation, galaxies mostly form from the cooled gas inside dark matter substructures in haloes. Thus, instead of arbitrary positions inside haloes, we are more interested in the substrutures (or subhaloes), where galaxies and stars reside such that tests of gravity are possible, for example, by studying the effect of modified gravity on stellar evolution and properties \citep{Davis2012,bvj2013}. Since subhaloes represent small density peaks inside a halo, with densities higher than their immediate surroundings, it is reasonable to expect the chameleon screening inside them to be stronger than outside. Many modified gravity simulations, however, do not have sufficient resolution to resolve subhaloes. As a result, \citet{ctl2015} propose to approximate the fifth force ratio inside a subhalo, which is at a distance $r$ from the centre of its host halo, to be the average value at all dark matter simulation particles inside a thin shell with radius range $[r-\Delta r/2,r+\Delta r/2]$. The {\sc liminality} simulation has high enough resolution for us to check the accuracy of this approximation. We apply a minimum 40 particles subhalo mass cut in order to reduce numerical noise.

\begin{figure*}
\includegraphics[width=\textwidth]{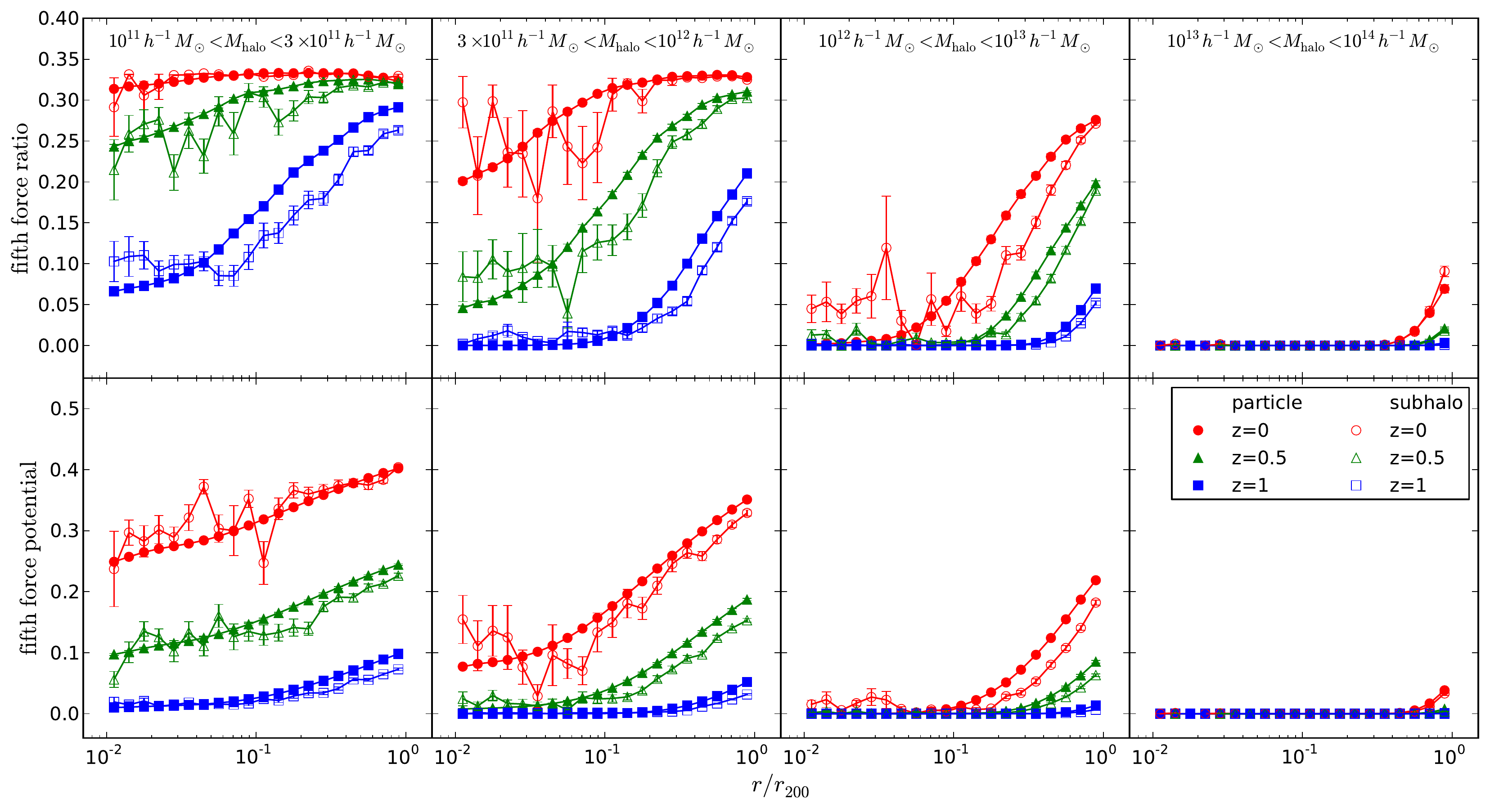}
\caption{The stacked fifth force ratio (upper panels) and fifth force potential (bottom panels) profiles within halo radius $R_{200}$ at $z=0$ (red circles), $0.5$ (green triangles) and $1$ (blue squares). In each panel we show the comparison between the quantities measured by all simulation particles (filled symbols) and that measured by subhaloes (open symbols).}
\label{fig:sub}
\end{figure*}

In Fig.~\ref{fig:sub}, we show the fifth force ratio and fifth force potential as a function of the radial distance from the halo centre ($r/R_{200}$), for subhaloes (open symbols) and dark matter particles (filled symbols). We can see that inside the most massive haloes (the rightmost column) and the least massive haloes (the leftmost column), the degree of screening is similar for dark matter particles and subhaloes. For the former, both particles and subhaloes are perfectly screened, which is why there is little difference; for the latter, at low redshifts, both particles and subhaloes are unscreened, such that again there is no difference.

For haloes of intermediate masses (the middle columns), the screening is consistently stronger in subhaloes where densities are higher than their immediately surroundings, as expected. However, the difference is generally small, because the Newtonian potential inside subhaloes is not dramatically deeper than outside. Although the curves for subhaloes are noisier due to poor statistics, they follow the trend of the curves for dark matter particles, which suggests that the approximation of \citet{ctl2015} can provide a reasonable conserative estimate of screening in subhaloes even for lower-resolution simulations where subhaloes are unresolved.

Fig~\ref{fig:sub} (filled symbols) also helps to visualize the `screening profile' inside dark matter haloes. We can see how the inner parts of haloes are completely screened for massive haloes, while more and more volume becomes unscreened for smaller haloes and at lower redshifts. In particular, we note that the transition from screened to unscreened regions is quite slow.

\section{Discussion and conclusions}

\label{sect:conclusion}

We have investigated the effect of environment on the efficiency of chameleon screening in $f(R)$ gravity. Based on a high-resolution $N$-body simulation \citep{liminality}, we have checked the various ways to define the `environment' of a dark matter halo. The definitions can be roughly put into three categories:

1. counting how many galaxies or, in $N$-body simulations, haloes, a given halo has as neighbours which satisfy certain requirements on their mass and/or distance from the considered halo;

2. estimating the underlying (nonlinear) dark matter density given the halo/galaxy number density;

3. using the Newtonian potential caused by the matter density field at the positions and surroundings of a given halo.

Each of these classes of environment definitions can be further divided depending on the precise physics included and parameters used, and the resulting definitions are given in Table \ref{table:environments}. In Fig.~\ref{fig:corr} we show the correlations of the different environment measures, where we find overall a good agreement among them.

We then study how the screening of the fifth force inside dark matter haloes depends on the environment that these haloes live in. Our analysis reconfirms the well known result that the screening is stronger for more massive haloes that live in dense environments. More importantly, the result also shows a reasonable agreement between the different environment definitions, hence verifying the robustness of the latter. This will have important implications for the construction of `screening maps' from observed galaxy catalogues \citep[see, e.g.,][]{bvj2013}, which is an essential step for predicting precisely how gravity changes its behaviour inside galaxies, which in turn can be used to constrain any deviations from GR. Since the model we study, F6, deviates only slightly from GR, being able to confidently rule it out will push the boundary of cosmological tests of gravity firmly into a new regime.

These different environment measures require different analyses of observation data: the conditional nearest neighbour measure can be directly applied to observed galaxy catalogues, the spherical overdensity measure requires a reconstruction of the matter density field from the observed galaxy field, while the experienced gravity measure requires a derivation of the three-dimensional Newtonian (and lensing) potential, which can be obtained by using weak lensing tomography. Because systematical errors in these analyses could lead to mis-identification of the environment, one can combine the different environment definitions if observational data allow.

We have also considered the screening of the fifth force inside dark matter subhaloes, and confirmed that the screening is stronger than in their host main haloes on average, as subhaloes have higher densities than their surroundings. However, the difference is small and the fifth force at the positions of simulation particles can act as a reasonable upper bound of subhaloes at the same positions. This result is useful, since it means that lower-resolution simulations of chameleon $f(R)$ gravity, even though unable to accurately resolve subhaloes, can still provide useful information about how well the fifth force is screened inside subhaloes and the galaxies in them.

Using the results here, we will be able to make screening maps of the Universe. This will be left for a future work.

\section*{Acknowledgements}
The authors wish to thank Carlton Baugh for helpful discussion. This work was supported by Science and Technology Facilities Council (ST/L00075X/1) and Cosmic Visions Euclid Science Support (ST/P006299/1). This work used the DiRAC Data Centric System at Durham University, operated by the Institute for Computational Cosmology on behalf of the STFC DiRAC HPC Facility ({http://www.dirac.ac.uk}). This equipment was funded by BIS National E-infrastructure capital grant ST/K00042X/1, STFC capital grant ST/H008519/1, STFC DiRAC Operations grant ST/K003267/1 and Durham University. DiRAC is part of the National E-infrastructure.







\appendix

\section{Consistency check of the $\Phi_\star$ environment measure}
\label{sect:appendix_a}

\begin{figure*}
\includegraphics[width=\textwidth]{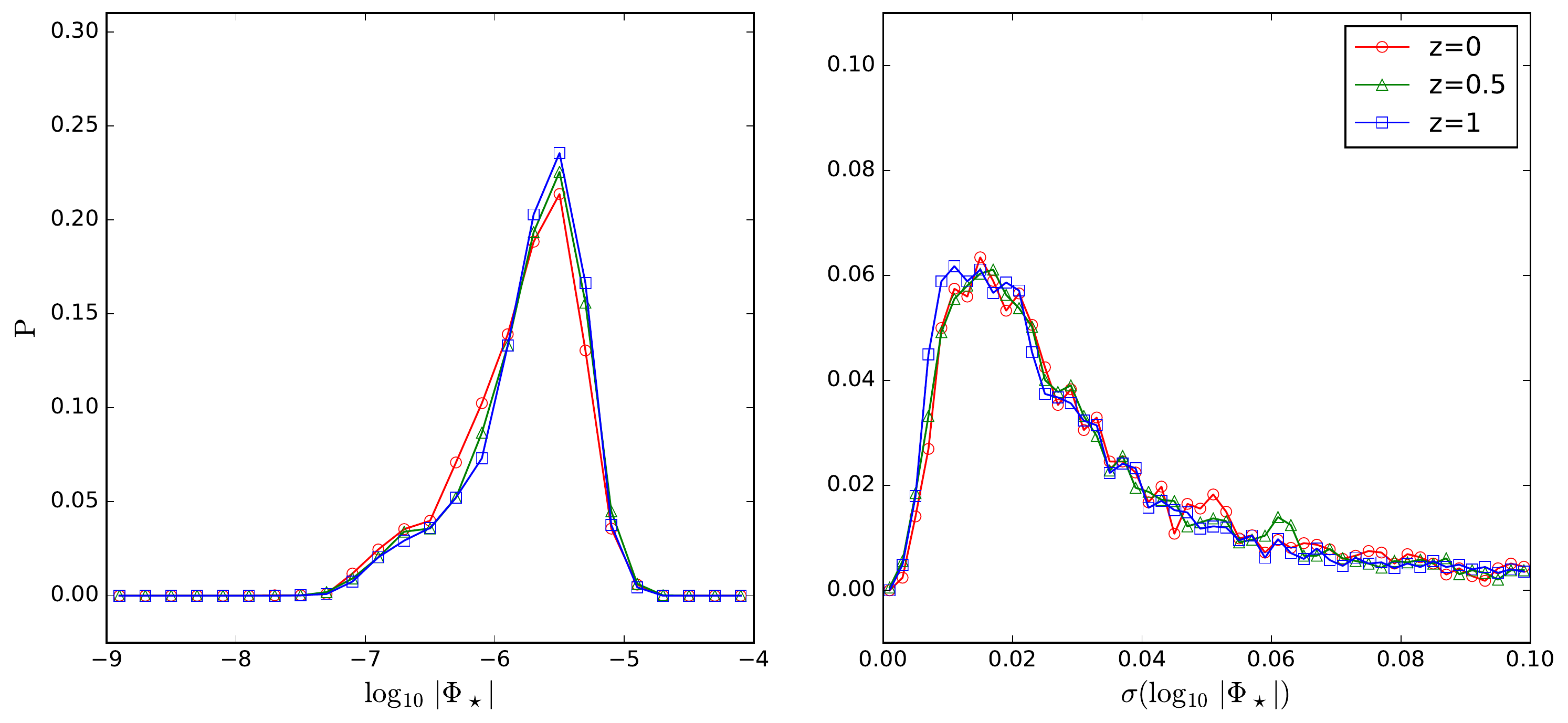}
\caption{The probability distribution of the average value of $\Phi_\star$ inside dark matter haloes (left panel) and the standard deviation of $\log_{10}|\Phi_\star|$ across 20 different radius bins (right panel). The mass range of the haloes is $1\sim 3 \times 10^{11} \msun$. The variation of $\Phi_\star$ is very small inside haloes, confirming that $\Phi_\star$ is mainly determined by the larger-scale structures, and therefore is a good environment measure.}
\label{fig:phi_inf_dist}
\end{figure*}

While defining the experienced gravity environment measure, $\Phi_\star$, we have subtracted the self Newtonian potential due to a halo itself from the total potential inside the halo, so that $\Phi_\star$ is created by all matter outside the halo, including the large-scale structure.
Since the size of the halo is generally negligible compared with the effective size of surrounding environment, we would expect $\Phi_\star$ to vary little inside it. As a result, an important consistency check is to check that the $\Phi_\star$ numerically obtained from the simulation does indeed have very little fluctuation inside haloes, for example, across the different bins of radial distance.

For this check, we divide the halo radius, $r_{200}$, into 20 bins equally spaced in logarithmic scale and calculate the values of $\Phi_\star(r)$ in each spherical shell according to Eq.~(\ref{eq:Phi}). The value of $\Phi_\star$ for a halo quoted in this paper is the average of these 20 values. The left panel of Fig.~\ref{fig:phi_inf_dist} shows the distribution of this average value for all haloes in the lowest halo mass bin (just for example). We see that the environment potential of haloes $\Phi_\star$ peaks at around $|\Phi_\star|=10^{-5.5}$, and is between $10^{-7}\sim10^{-5}$ for most haloes. There is very little redshift evolution of this distribution.

We also calculate the standard deviation of the $\log |\Phi_\star|$ values in all bins for each halo. The right panel displays the distribution of standard deviation of $\log_{10}|\Phi_\star|$ for the same haloes as considered in the left panel. Although the values of $\log_{10}|\Phi_\star|$ mostly fall between $-7$ and $-5$, the variation in the different radius bins is fairly small, peaking at $\sim0.02$ and smaller than $0.1$ for most haloes. This indeed confirms that $\Phi_\star$ is almost a constant within dark matter haloes, and not affected by the fairly strong dependence of the total gravitational potential $\Phi(r)$ on the radial distance $r$.


\bsp	
\label{lastpage}
\end{document}